# Measurement of the Charge Transfer Efficiency of Electrons Clocked on Superfluid Helium

G. Sabouret and S. A. Lyon

*Department of Electrical Engineering, Princeton University, Princeton, New Jersey 08544*

**Abstract:** Electrons floating on the surface of liquid helium are possible qubits for quantum information processing. Varying electric potentials do not modify spin states, which allows their transport on helium using a charge-coupled device (CCD)-like array of underlying gates. This scheme depends upon efficient inter-gate electron transfer and on the absence of electron traps. We will present a measurement of the charge transfer efficiency (CTE) of electrons clocked back and forth above a short CCD-like structure. The CTE obtained at low clocking frequencies is 0.999 with an electron density of about 4 electrons/µm$^2$. We find no evidence for deep electron trapping.

Electrons floating on the surface of superfluid helium are potential candidates as qubits for quantum computation[1,2] relying either on their charge states[3,4,5] or spin states[6] as two-level quantum systems. In the case of spin qubits, the electrons would be moved about the surface of the helium using a charge coupled device (CCD)-like network of underlying gates, making this scheme intrinsically scalable. A well-known concern for silicon CCD's is their charge transfer efficiency (CTE) due to charge trapping at the Si/SiO$_2$ interface or in bulk traps[7]. Whereas leaving a few electrons behind on the clocking path is tolerable in silicon CCD's due to the large number of electrons involved, it becomes unacceptable for quantum computing where single electrons reside on each gate[8] and conserving the electrons is critical.

Here we report the measurement of the CTE for electrons on superfluid helium in a short CCD. The mobility of electrons on liquid helium has traditionally been measured[9,10] using three submerged gates. With the central gate gounded, an alternating current (ac) potential applied to one of the outer gates makes the electrons move back and forth on the surface. The voltage that they induce on the second outer gate is partially out of phase with the driving ac signal and the electron mobility can be computed from the phase shift. These experiments, however, only measure the mobility of electrons that are free to move. Any charge staying trapped for a time longer than the period of the applied ac potential does not contribute to the signal and therefore goes undetected, making CTE prediction impossible from these experiments. In this paper we present a direct measurement of the CTE using a seven-gate CCD-like structure. The gates are used to clock electrons back and forth in a way similar to the conventional three-plate experiments. The difference lies in the fact that we apply a special gate voltage sequence that expels electrons unable to transfer away from a chosen gate on each cycle. We find evidence of weak trapping probably caused by gate roughness. However, if electrons are given enough time to diffuse, we measure a CTE of 0.999 with an electron density of $2.6 \times 10^7$ $cm^{-2}$.

Figure 1 is a diagram of our experimental setup. A hermetically sealed copper cell is held at a temperature of 1.55K in a pumped $^4$He system. Stycast feedthroughs allow electrical connections to the inside of the cell. A mass flowmeter is used to determine the quantity of $^4$He gas admitted and therefore sets the level of liquid helium inside the cell. That cell also contains our gate structure at the bottom, 1.17 mm away from a ground plate at the top (both made of copper clad printed circuit board). The gates and ground plate form a parallel plate capacitor that determines the density of the electrons on the surface of the liquid helium with a constant (dc) positive 4 V bias voltage applied to all the bottom gates. A filament lies just above an opening in the top ground plate and is briefly pulsed at the beginning of the experiment to generate free electrons. A guard ring (negatively biased at −4 V) surrounds the seven bottom gates to isolate them from electrons in the rest of the cell. Besides the dc voltage bias common to all seven gates, independent alternating current (ac) signals can be applied to gates #2 to 7. Gate #1 was used for detection and connected to our lock-in amplifier. The helium level in our experiments was set to be 0.9 mm above the bottom gates.

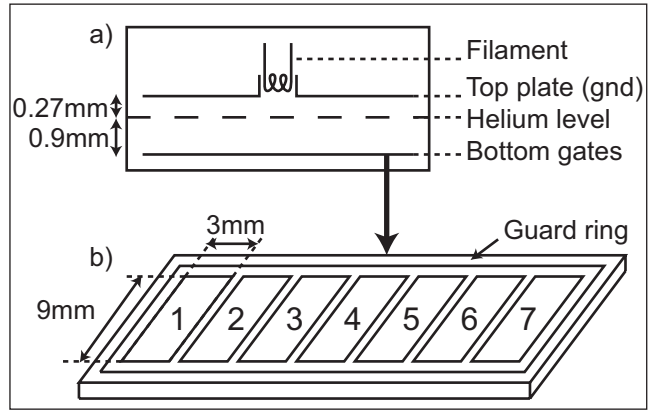

FIG. 1. a) Schematic of experimental cell seen from the side. The top plate is grounded while a common dc bias is applied to all the bottom gates to hold the electrons on the helium surface. b) View of the seven bottom gates and the guard ring. The helium level lies 0.9 mm above them. Gate #1 is used as a detection gate and gate #7 is used to eject the electrons.

We first ran a conventional three-phase CCD gate voltage sequence (alternating between driving the electrons to the left and right) on gates #3 through 7 to move electrons back and forth between gate #1 and gate #7. The potential on gate #2 was kept constant (or varied slowly) so that it acted as an ac ground to shield gate #1 from any signal coupled from the changing voltages on the other gates. The clocking sequence moved electrons towards gate #7 for three sub-periods (fig. 2i), paused for three sub-periods (fig. 2ii), moved the electrons towards gate #1 for three sub-periods (fig. 2iii), paused again for three sub-periods (fig. 2ii) and then repeated. The reference frequency of the lock-in amplifier was set to be equivalent to a round trip cycle of the electrons (12 sub-periods). Pulsing the filament under these conditions created a signal on the lock-in amplifier and that signal remained stable for the entire time of our observations (hours) at lock-in frequencies as high as 100 kHz (100 000 round trips per second). This high stability shows that no electrons were crossing the guard ring. Changing the dc voltage on gate #2 to 0 V caused the signal to disappear, since electrons were clocked onto pixel #1, but could not cross that barrier back to the other pixels. Bringing gate #2 back to +4 V allowed us to completely recover the signal, showing again its stability and demonstrating that it arose from clocking electrons rather than from some capacitive coupling of the ac gate voltages onto the detection gate.

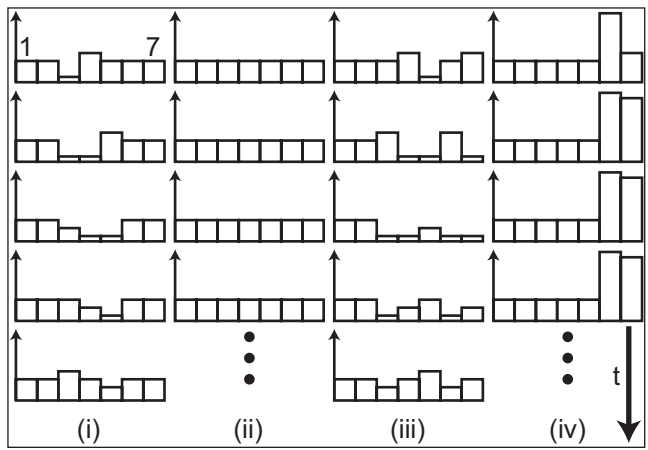

FIG. 2. Diagram of the negative of the individual ac voltages applied to each of the seven gates numbered from left to right. Gate #1 remains at a constant holding voltage of +4 V while the other gates range from 3 V more attractive to 9 V more repulsive. The potentials are drawn to scale. For stably shifting the electrons back and forth, the first sub-period (i) is repeated three times and clocks the electrons from gate #1 to gate #7, then the system pauses for three sub-periods (ii) and then clocks the electrons back from gate #7 to gate #1 for three sub-periods (iii) after which it pauses again (ii). For the CTE measurements, the first pause (ii) is replaced by the modified sequence (iv) that expels all electrons remaining on gate #7.

The density of the electrons on the surface of the helium can be computed in two ways. The first one considers the seven bottom gates and the top ground gate to be a parallel plate capacitor. Electrons will accumulate on the surface until their self-induced potential compensates the applied dc voltage[11]. The density is therefore given by $n_s=eV_0/de$ where $e$ is the dielectric constant of helium, $V_0$ is the applied dc voltage, $e$ is the electron charge and $d$ is the thickness of the helium. The electron density computed in this manner was $n_s=2.6 \times 10^7$ cm$^{-2}$ (or about one electron per 4 µm$^2$). Another way to estimate $n_s$ is to look at the signal amplitude on the lock-in amplifier and at the capacitance of the lock-in input (including the wire connecting the amplifier input to the detection gate). In this case $n_s=CV_s h/Aed$ where $C \sim 400$ pf is the capacitance of the wire and the lock-in input, $V_s \sim 400$ mV is our signal, including compensation for losses in the lock-in filter stage, $A = 0.285$ cm$^2$ is the area of the detection gate and $h$ is the separation between the bottom plates and the top ground plate. While there are considerably more uncertainties using this method as compared to the first (the lock-in signal is sensitive to the shape of the detected waveform, for example) it gives us $n_s = 1.5 \times 10^7$ cm$^{-2}$. The reasonable agreement between these two numbers is further evidence that the electrons actually do travel through the pixels.

The CTE measurement was made by first running the previous sequence to allow any excess electrons that would have been expelled by the changing gate potentials to escape to the top and walls of the cell and by then modifying the gate pulse sequence as follows: the three sub-periods following clocking from gate #7 to detection gate #1 were modified so that gate #6 was biased to –5V (a higher barrier than the guard ring) to isolate any electrons left on gate #7 (fig. 2iv). The potential on gate #7 was then lowered to –4V, expelling any leftover electrons from the system towards the top plate and the walls of the cell – all kept at ground.

We expect the electrons to move from one gate to the next through diffusion, since the field lines are mostly vertical in this system due to the large size of the gates. As the electrons diffuse from one gate to another a constant fraction will remain behind. The fraction left on gate #7 is expelled on every cycle, so we expect an exponential decay of the measured signal. This is indeed what we observed for measurements taken at 7 Hz and 13 Hz (figure 3, final part of the decay). Fitting the exponential decay as a function of the number of cycles allowed us to extract the CTE. It was found to be 0.9990 at 7 Hz and 0.9974 at 13 Hz. A lower CTE for a higher frequency is expected as electrons have less time per cycle to diffuse from one gate to another resulting in more electrons left on gate #7 and expelled. However we found that for frequencies of 23 Hz and above, the decay was linear instead of being exponential, meaning that a fixed number of electrons was expelled on each cycle (see the inset of figure 3 for a typical trace above 23 Hz). We believe that this is evidence for very shallow trapping due to the fact that our bottom gates are not perfectly smooth or leveled; the local potential well therefore varies. Allowing more helium inside the cell in order to slightly raise the liquid level was found to slow down the linear decay as distance smoothes out the uneven potential.

The first part of the decay, which is exponential on every trace, remains unexplained. The fact that the transition between the initial and the final decays creates a discontinuity in the slope of the signal would suggest a rapid transition that happens upon reaching some constant critical density. Further investigation is needed to understand this effect.

We argued above that electron diffusion is the dominant process determining the decay at low frequencies. This can be checked by solving the one-dimensional diffusion equation for gate #7. The electron density at the boundary with gate #6 vanishes whenever the potential on gate #6 is more attractive than the potential on gate #7 (the electrons on gate #6 cannot return to interact with the electrons on gate #7; any electron coming from gate #7 therefore sees an effective zero electron density at the boundary). The other boundary was set to be a hard potential wall (density gradient is zero). The diffusivity leading to the observed signal decay was found to be $D$=17 cm$^2$/s using this technique. From Einstein's relation the mobility is found to be m=1.27x10$^5$ cm$^2$/Vs which agrees with published experimental data obtained in other ways[9].

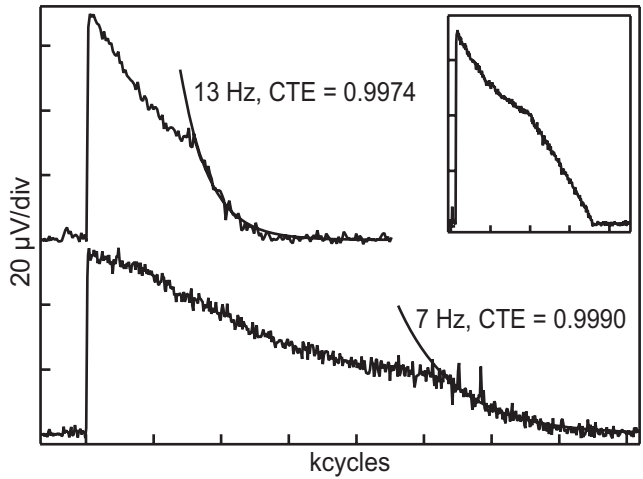

FIG. 3. Decaying signals obtained using the modified sequence that expels electrons from gate #7 at two lock-in frequencies (the frequency of a round trip of the electron above the gates). We fit the final part of the decay to obtain the CTE. The inset represents a typical signal at frequencies of 23 Hz and higher: the final part of the decay is linear instead of being exponential. There is an initial exponential decay at all frequencies which appears to saturate and whose origin is not currently understood.

In this experiment, we have demonstrated clocked transfer of electrons on the surface of liquid helium using a short CCD-like geometry. By modifying the clock sequence, we are able to measure the charge transfer efficiency and show that it can be over 0.999. This CTE was measured for low clocking frequencies since the mechanism governing electron transfer from one gate to another is diffusion and our gates are large (3 mm width). We expect much faster operating frequencies with narrow gates where the fringing fields help the electron transfer as in a conventional buried-channel CCD. It is remarkable that we are able to get such a high CTE at such low electron densities. By way of comparison, silicon CCD's normally operate at electron densities four or five orders of magnitude higher[12]. In order to operate at lower densities, silicon CCD's often need "fat zero's" to fill electron traps and get a good CTE[12,13]. We have shown that the density of deep traps in our system is exceptionally low. We do however see shallow trapping, probably due to non-uniform holding fields that will become irrelevant for small structures with substantial fringing fields for electron transfer. The apparent absence of deep traps and the fact that the CTE is not reduced at a low electron density (comparable to the density required for mobile electron-spin qubit quantum computing) is encouraging.

This work was supported in part by the NSF under grant CCF-0323472, and by the ARO and ARDA under contract W911NF-04-1-0398.